\documentclass[pra,aps]{revtex4}
\usepackage{graphicx}
\usepackage{bm}

\newcommand{\ket}[1]{| #1 \rangle}

\newcommand{\be}{\begin{equation}}
\newcommand{\ee}{\end{equation}}
\newcommand{\bea}{\begin{eqnarray}}
\newcommand{\eea}{\end{eqnarray}}
\newcommand{\beas}{\begin{eqnarray*}}
\newcommand{\eeas}{\end{eqnarray*}}
\newcommand{\bes}{\begin{equation*}}
\newcommand{\ees}{\end{equation*}}

\newcommand{\Aeff}{A_{\rm eff}}

\begin{document}

\title{Quantum theory of degenerate $\chi^{(3)}$ two-photon state}
\author{Jun Chen, Kim Fook Lee, and Prem Kumar}
\address{Center for Photonic Communication and Computing, EECS Department\\
Northwestern University, 2145 Sheridan Road, Evanston, IL
60208-3118}

\maketitle

We develop a theory to model the degenerate two-photon state
generated by the 50/50 Sagnac-loop source. We start with an
interaction Hamiltonian that is capable of describing the
interaction among the four optical fields (non-degenerate pump and
degenerate signal/idler), which reads: \bea H_I(t) &=& \alpha \int
dV (E_s^{(-)} E_i^{(-)} E_{p1}^{(+)} E_{p2}^{(+)} + {\rm H.c.})
\label{HRDFWM} \eea where $\alpha$ is a material constant that is
characteristic of the optical fiber being used. The non-degenerate
pump field is taken to be two synchronous pulses (denoted by
subscripts $p1$ and $p2$), copolarized and co-propagating down the
fiber axis (denoted as {\it z} direction here), with central
frequencies $\Omega_{p1}$ and $\Omega_{p2}$ and envelope shapes
$\overline{E}_{p1}$ and $\overline{E}_{p2}$. Mathematically, they
are written as below \bea E_{p1}^{(+)} &=& \int d\omega_{p1}
\overline{E}_{p1}(\omega_{p1})\,e^{i k(\omega_{p1}) z - i
\omega_{p1} t}\,e^{-i \gamma P_1 z}\,,
\label{EP1-form} \\
E_{p2}^{(+)} &=& \int d\omega_{p2}
\overline{E}_{p2}(\omega_{p2})\,e^{i k(\omega_{p2}) z - i
\omega_{p2} t}\,e^{-i \gamma P_2 z}\,, \label{EP2-form} \eea where
$\omega_{p1}=\Omega_{p1}+\nu_p$ and
$\omega_{p2}=\Omega_{p2}+\nu'_p$ are the frequency arguments for
the two pump fields. $\nu_p$ ($\nu'_p$) denotes the frequency
component within $p1$'s ($p2$'s) spectrum that deviates from its
central frequency $\Omega_{p1}$ ($\Omega_{p2}$) by that amount.
The phase tags $e^{-i \gamma P_1 z}$ and $e^{-i \gamma P_2 z}$ are
induced by $p1$'s and $p2$'s self-phase modulation (SPM)
respectively, and are included in a straightforward manner. Note
that $P_1$ and $P_2$ denote the peak powers of $p1$ and $p2$,
respectively.

\begin{figure}
\centerline{\includegraphics[scale=0.5]{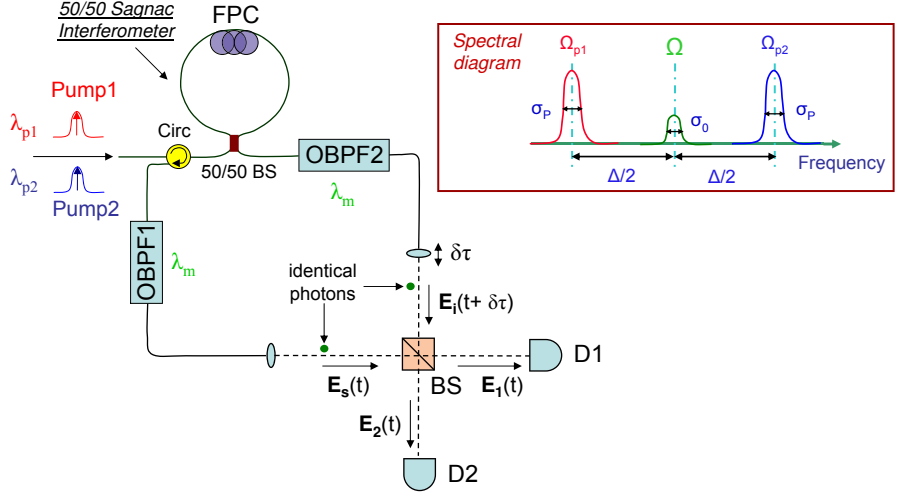}}
\caption{Schematic of the Hong-Ou-Mandel experiment with the 50/50
Sagnac-loop identical photon source. FPC, fibre polarization
controller; BS, beam splitter; OBPF, optical bandpass filter; D1,
D2, photon-counting detectors. Inset shows the spectral diagram of
the non-degenerate pump and degenerate signal/idler fields.
$\Omega_{p1}$, pump-1 ($p1$) central frequency; $\Omega_{p2}$,
pump-2 ($p2$) central frequency; $\Omega$, signal/idler central
frequency; $\sigma_p$, pump bandwidth; $\sigma_0$, OBPF bandwidth;
$\Delta/2$, central frequency difference between P2 and
signal/idler (or signal/idler and P1); $\mathbf{E_s}$,
$\mathbf{E_i}$, $\mathbf{E_1}$ and  $\mathbf{E_2}$, electrical
fields before and after the BS, see text for details.}
\label{DTPS1}
\end{figure}

The degenerate signal/idler field, with a center frequency at
$\Omega$, is quantized according to~\cite{Grice97}: \bea E_s^{(-)}
&=& \int d\omega_s\, A(\omega_s)\, a_s^{\dag} \,e^{-i[k(\omega_s)z
-\omega_s t]} \,,\label{ES-form} \\
E_i^{(-)} &=& \int d\omega_i\, A(\omega_i)\, a_i^{\dag}\,
e^{-i[k(\omega_i)z - \omega_i t]}\,,\label{EI-form} \eea where
$a_j^{\dag}\,(j=s,i)$ is the creation operator for the $j$ mode
with frequency $\omega_j$ and wave-vector magnitude
$k(\omega_j)=\displaystyle\frac{n(\omega_j)\,\omega_j}{c}$.
$\omega_s=\Omega+\nu_s$ and $\omega_i=\Omega+\nu_i$ represent the
frequency of signal and idler photon, respectively, where $\nu_s$
and $\nu_i$ are the deviations for each photon's frequency from
their central frequency $\Omega$.
$A(\omega_j)=-i\sqrt{\displaystyle\frac{\hbar\omega_j}{2
\epsilon_0 n^2(\omega_j)}}$ is a slowly varying function of
frequency and may be taken outside the integral. Now the
interaction Hamiltonian may be expressed as \bea H_I(t) &=& A
\int_{-L}^{0} dz \int d\omega_s \int d\omega_i\,
a_s^{\dag}\,a_i^{\dag} \int d\omega_{p1} \int d\omega_{p2}
\,e^{-i\gamma (P_1 + P_2) z}\,\overline{E}_{p1}(\omega_{p1})\,\overline{E}_{p2}(\omega_{p2}) \nonumber\\
&&{\rm exp}\{i
[k(\omega_{p1})+k(\omega_{p2})-k(\omega_{s})-k(\omega_{i})] z - i
(\omega_{p1}+\omega_{p2}-\omega_s-\omega_i) t  \}\,, \label{HI2}
\eea where $L$ is the length of the fiber, and $A$ is an overall
constant consisting of $\alpha$, $A(\omega_j)$ and the effective
cross-section of the fiber $\Aeff$.

The state vector at the output of the fiber can be calculated by
means of first-order perturbation theory, namely, \begin{eqnarray}
|\Psi\rangle_{\rm out} &=& |0\rangle +
  \frac{1}{i\,\hbar}\int_{-\infty}^{\infty} H_I(t)\, dt
  \,|0\rangle\,,
\end{eqnarray} where the second term is the two-photon state that
we seek, which we denote simply by $\ket{\Psi}$. After taking into
account the fact that \bea \int_{-\infty}^{\infty} e^{-i
(\omega_{p1}+\omega_{p2}-\omega_s-\omega_i) t} dt &=& 2\pi
\delta(\omega_{p1}+\omega_{p2}-\omega_s-\omega_i)\,,
\label{deltafunction} \eea we arrive at the following expression
of the two-photon state: \bea \ket{\Psi} &=& \frac{2\pi A}{i
\hbar}\,\int_{-L}^{0} dz \int d\omega_s \int d\omega_i\,
a_s^{\dag}\,a_i^{\dag} |0\rangle\,\int d\omega_{p1}\,e^{-i\gamma
(P_1 + P_2) z} \nonumber\\
&&\overline{E}_{p1}(\omega_{p1})\,\overline{E}_{p2}(\omega_{s}+\omega_i
-\omega_{p1})\,e^{i [k(\omega_{p1})+k(\omega_{s}+\omega_i
-\omega_{p1})-k(\omega_{s})-k(\omega_{i})] z} \label{DTPS-fun}
\eea where we have utilized the $\delta$-function to simplify the
integral over $\omega_{p2}$, which also reinforces the energy
conservation among the four interacting optical fields.

To further simplify our analysis, we refer to the inset ``Spectral
diagram" shown in Fig.~\ref{DTPS1}, which clearly illustrates the
various system parameters by their corresponding mathematical
symbols. These parameters correspond to the experimental settings
schematically depicted in the main part of Fig.~\ref{DTPS1}. The
two pump pulses are both assumed to be Gaussian-shaped with equal
amplitude and equal bandwidth, i.e., \bea
\overline{E}_{p1}(\omega) &=& E_p \,
e^{-\frac{(\omega-\Omega_{p1})^2}{2 \sigma_p^2}}\,, \nonumber\\
\overline{E}_{p2}(\omega) &=& E_p \,
e^{-\frac{(\omega-\Omega_{p2})^2}{2 \sigma_p^2}}\,,
\label{PumpShape} \eea where $P_1=P_2 \equiv P_p \propto
E_p^2\,\sigma_p^2$ are the peak powers of the two pump pulses, and
$\sigma_p$ denotes their common optical bandwidth. By using Taylor
expansion at the frequency $\Omega$ for the various $k$'s, we
obtain \bea \Delta k &\equiv& k(\omega_{p1}) + k(\omega_s +
\omega_i -\omega_{p1}) -
k(\omega_s) - k(\omega_i) \nonumber\\
&=& k''(\Omega) \left[\left(\frac{\Delta}{2}-\nu_p\right)^2 +
\left(\frac{\Delta}{2}-\nu_p\right)\left(\nu_s+\nu_i\right) +
\nu_s\,\nu_i\right]\,, \label{kvector} \eea where we keep the
expansion series to second-order dispersion only, which proves to
be sufficient in most cases.

The $\omega_{p1}$-integral in Eq.~(\ref{DTPS-fun}) can be
rewritten as \bea \Phi(\nu_s, \nu_i, z) &\equiv& \int d\omega_{p1}
\overline{E}_{p1}(\omega_{p1})\, \overline{E}_{p2}(\omega_s +
\omega_i - \omega_{p1})\,e^{i \Delta k z} \nonumber\\
&=& E_p^2 \int d\nu_p \,e^{-\frac{\nu_p^2 +
(\nu_s+\nu_i-\nu_p)^2}{2\sigma_p^2} + i \beta_2 z
\left[\left(\frac{\Delta}{2}-\nu_p\right)^2 +
\left(\frac{\Delta}{2}-\nu_p\right)\left(\nu_s+\nu_i\right) +
\nu_s\,\nu_i \right]}\,, \label{Phi-fun} \eea where in the last
step we have used Eqs.~(\ref{PumpShape}) and (\ref{kvector}), and
$\beta_2$ is a shorthand for $k''(\Omega)$. We are then left with
the length integral \bea Q(\nu_s, \nu_i) &\equiv& \int_{-L}^{0} dz
\,\Phi(\nu_s, \nu_i, z)\,e^{-2 i \gamma P_p z}\,,\label{Q-fun}
\eea and the two-photon state in Eq.~(\ref{DTPS-fun}) is
reorganized into \bea \ket{\Psi} &=& \frac{2 \pi A}{i \hbar} \int
d\nu_s \int d\nu_i\,Q(\nu_s, \nu_i) \ket{\Omega + \nu_s}
\ket{\Omega + \nu_i}\,,\label{DTPS-fun2} \eea where $\ket{\omega}$
is a one-photon Fock state populated with a single photon of
frequency $\omega$. We remark that the function $Q(\nu_s, \nu_i)$
is completely analogous to $\alpha(\omega_o
+\omega_e)\,\Phi(\omega_o, \omega_e)$ in Eq. (9) in
Ref.~\cite{Grice97}. Similarly, $|Q(\nu_s, \nu_i)|^2$ can be
interpreted as the probability distribution of the two-photon
state~\cite{Grice97}. However, the apparent symmetry of
$\Phi(\nu_s, \nu_i, z)$, and thus $Q(\nu_s,\nu_i)$, with respect
to its two frequency arguments results in qualitatively different
behavior for the two-photon state from that in
Ref.~\cite{Grice97}, which is asymmetric in its frequency
arguments.

Further evaluation of $\Phi(\nu_s, \nu_i, z)$ is made possible by
using the integral formula from Ref.~\cite{Gradshteyn}, which
deals with Gaussian integrals with complex arguments, resulting in
\bea \Phi(\nu_s, \nu_i, z) &=& \sqrt{\pi}\, \sigma_p
\,E_p^2\,e^{-\frac{(\nu_s + \nu_i)^2}{4 \sigma_p^2}}\,\frac{{\rm
exp}\left[-\frac{\beta_2^2\,z^2\,\Delta^2\,\sigma_p^2}{4(1+\beta_2^2\,z^2\,\sigma_p^4)}\right]}
{\sqrt[4]{1+\beta_2^2\,z^2\,\sigma_p^4}}\,
{\rm exp}\left\{\frac{i \beta_2 z}{4}\left[\Delta^2-(\nu_s-\nu_i)^2\right]\right\} \nonumber\\
&& {\rm exp}\left[\frac{i}{2}\,\arctan(\beta_2\, z \,\sigma_p^2) -
i\frac{\beta_2^3\,z^3\,\Delta^2\,\sigma_p^4}{4(1+\beta_2^2\,z^2\,\sigma_p^4)}
\right]\,, \label{Phi-fun2} \eea where $\Delta \equiv
\Omega_{p2}-\Omega_{p1}$ is the central frequency difference
between the two pump fields. We have thus formally obtained the
expression for the two-photon state, which is given by
Eq.~(\ref{DTPS-fun2}) or its following alternative version: \bea
\ket{\Psi} &=& \frac{2 \pi A}{i \hbar} \int d\omega_s \int
d\omega_i\,\widetilde{Q}(\omega_s, \omega_i) \ket{\omega_s}
\ket{\omega_i}\,,\label{DTPS-fun3} \eea where
$\widetilde{Q}(\omega_s, \omega_i)$ is equivalent to
$Q(\omega_s-\Omega, \omega_i-\Omega)$ given by Eq.~(\ref{Q-fun}).

After obtaining the two-photon state, we are now ready to analyze
the experiment shown schematically in Fig.~\ref{DTPS1}. A variable
delay $\delta \tau$ is inserted in one photon's path, before the
two identical photons are recombined at the 50/50 beam-splitter
(BS). As shown in Fig.~\ref{DTPS1}, if we denote the
electric-field operators before the BS as $E_s^{(+)}(t)$ and
$E_i^{(+)}(t+\delta \tau)$, then the field operators after the BS
are given by \bea E_1^{(+)}(t) &=& \frac{1}{\sqrt{2}}
\left[E_s^{(+)}(t) + i E_i^{(+)}(t+\delta \tau)
\right] \,,\nonumber\\
E_2^{(+)}(t) &=& \frac{1}{\sqrt{2}} \left[i E_s^{(+)}(t) +
E_i^{(+)}(t+\delta \tau) \right]\,,\label{E1E2} \eea where the
vector nature of the field operators are ignored since they all
share the same polarization, and \bea E_{s,i}^{(+)}(t) \propto
\int d\omega_{s,i}\,a_{s,i}(\omega_{s,i})\,e^{-i \omega_{s,i}
t}\,e^{-\frac{(\omega_{s,i} - \Omega)^2}{2 \sigma_0^2}}
\label{EsEi} \eea are the electric-field operators before the BS
that include the shape of the OBPF, which is assumed to be
Gaussian here. The coincidence-count rate registered by detectors
D1 and D2 is given by~\cite{Glauber} \bea R_c(\delta \tau) &=&
\int_0^{\infty} dt_1 \int_0^{\infty} dt_2 \,P_{12}(t_1,t_2,\delta
\tau)\,, \label{CCrate} \eea where \bea P_{12}(t_1,t_2,\delta
\tau) &=& \langle\Psi|E_1^{(-)}(t_1) E_2^{(-)}(t_2) E_2^{(+)}(t_2)
E_1^{(+)}(t_1)|\Psi\rangle \nonumber\\
&=& |\langle 0|E_2^{(+)}(t_2) E_1^{(+)}(t_1)|\Psi\rangle|^2
\label{P12} \eea is the probability per pulse for coincidence
detection between the two detectors.

Eqs.~(\ref{DTPS-fun3}), (\ref{E1E2}), (\ref{EsEi}), and
(\ref{P12}), when plugged into Eq.~(\ref{CCrate}), after a simple
but lengthy calculation, yield \bea R_c(\delta \tau) &\propto&
\int d\omega_s \int d\omega_i \, |\widetilde{F}(\omega_s,
\omega_i)|^2\,\left[1-e^{-i (\omega_i-\omega_s) \delta \tau}
\right] \,,\nonumber\\
\widetilde{F}(\omega_s, \omega_i) &=& \widetilde{Q}(\omega_s,
\omega_i)\,{\rm exp}\left[-\frac{(\omega_s-\Omega)^2}{2
\sigma_0^2} - \frac{(\omega_i-\Omega)^2}{2 \sigma_0^2}\right]
\,,\label{CCrate2} \eea where we have used the fact that
$\widetilde{F}(\omega_s, \omega_i) = \widetilde{F}(\omega_i,
\omega_s)$ in obtaining the above results\footnote{We remark that,
in the case of $\widetilde{F}(\omega_s, \omega_i) \neq
\widetilde{F}(\omega_i, \omega_s)$, we would have obtained
$R_c(\delta \tau) \propto \int d\omega_s \int d\omega_i
\,\left[|\widetilde{F}(\omega_s, \omega_i)|^2 -
\widetilde{F}(\omega_s, \omega_i)\,\widetilde{F}^*(\omega_i,
\omega_s)\,e^{-i (\omega_i-\omega_s) \delta \tau} \right]$.}. Its
alternative version, written in terms of the difference-frequency
$\nu_s$ and $\nu_i$, reads \bea R_c(\delta \tau) &\propto& \int
d\nu_s \int d\nu_i \, |F(\nu_s, \nu_i)|^2\,\left[1-e^{-i
(\nu_i-\nu_s) \delta \tau}
\right] \,,\nonumber\\
F(\nu_s, \nu_i) &=& Q(\nu_s, \nu_i)\,{\rm exp}\left(-\frac{\nu_s^2
+ \nu_i^2}{2 \sigma_0^2} \right) \,.\label{CCrate3} \eea

It turns out that further simplification is possible for the above
Gaussian-filter case (by using again the formula from
Ref.~\cite{Gradshteyn}), which we explicitly write out as the
following: \bea R_c(\delta \tau) &\propto& \int_{-L}^0 dz_1
\int_{-L}^0 dz_2\,G(z_1)\,G^*(z_2)\,I(z_1, z_2) \,,\label{CCrate-G}\\
G(z) &=& \frac{{\rm
exp}\left[-\frac{\beta_2^2\,z^2\,\Delta^2\,\sigma_p^2}{4(1+\beta_2^2\,z^2\,\sigma_p^4)}\right]}
{\sqrt[4]{1+\beta_2^2\,z^2\,\sigma_p^4}}\, {\rm
exp}\left[\frac{i}{2}\,\arctan(\beta_2\, z\,
\sigma_p^2)\right]  \nonumber\\
&& {\rm exp}\left[ -
i\frac{\beta_2^3\,z^3\,\Delta^2\,\sigma_p^4}{4(1+\beta_2^2\,z^2\,\sigma_p^4)}\right]\,{\rm
exp}\left[i\frac{\beta_2\,\Delta^2\,z}{4} - 2 i \gamma P_p z
\right] \,,\label{GZ-fun} \\
I(z_1, z_2) &=&
\frac{\sqrt{2}\,\pi^2\,P_p^2\,\sigma_0^2}{\sigma_p\,\sqrt{\sigma_p^2
+ \sigma_0^2}}\,\frac{{\rm
exp}\left\{\frac{i}{2}\,\arctan\left[-\frac{\beta_2\,(z_1-z_2)\,\sigma_0^2}{2}\right]\right\}}
{\sqrt[4]{4+\beta_2^2\,(z_1-z_2)^2\,\sigma_0^4}} \nonumber\\
&& \left\{1-{\rm
exp}\left[-\frac{2\,\delta\tau^2\,\sigma_0^2}{4+\beta_2^2\,(z_1-z_2)^2\,\sigma_0^4}
 + i \frac{\beta_2\,(z_1-z_2)\,\delta\tau^2\,\sigma_0^4}{4+\beta_2^2\,(z_1-z_2)^2\,\sigma_0^4}
\right] \right\}\,. \label{Iz-fun} \eea

One can, for instance, generalize the above result to investigate
the effect that the filter has on the Hong-Ou-Mandel (HOM) dip.
Due to its experimental relevance, we shall write out explicitly
the formula that describes the case when the previously
Gaussian-shaped OBPFs are replaced with two identical
super-Gaussian filters, which reads \bea R_c(\delta \tau)
&\propto& \frac{\pi\,P_p^2}{\sigma_p^2} \int_{-L}^0 dz_1
\int_{-L}^0 dz_2 \int d\nu_s \int
d\nu_i\,G(z_1)\,G^*(z_2)\,e^{-\frac{(\nu_s+\nu_i)^2}{2
\sigma_p^2}}\,e^{-\frac{\nu_s^4+\nu_i^4}{\sigma_0^4}} \nonumber\\
&& {\rm exp}\left[-i\frac{\beta_2}{4} (\nu_s-\nu_i)^2(z_1-z_2)
\right]\,\left[1-e^{-i (\nu_i-\nu_s) \delta\tau} \right]
\,,\label{CCrate-SG} \eea where $G(z)$ is given by
Eq.~(\ref{GZ-fun}).

\begin{figure}
\centerline{\includegraphics[scale=0.5]{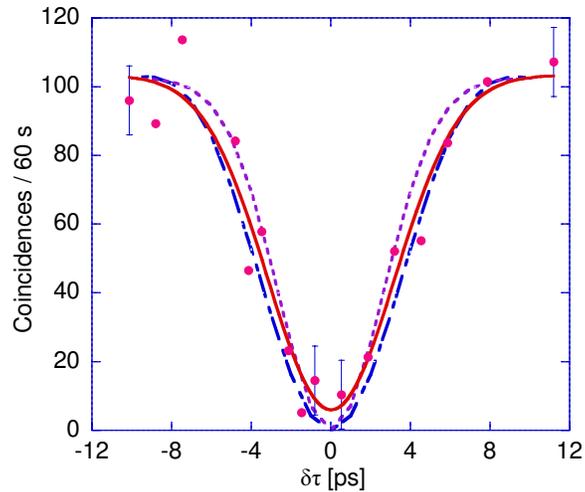}}
\caption{Theoretical predictions vs. experimental results. Pink
filled circles, experimental data; red solid curve, least-square
fitting for the data; purple dotted curve, theory fitting from
Eq.~(\ref{CCrate-G}) for Gaussian OBPFs; blue dot-dashed curve,
theory fitting from Eq.~(\ref{CCrate-SG}) for super-Gaussian
OBPFs. Realistic values for the experimental parameters used in
generating these curves are: $L=300$\,m, $\beta_2=-0.116\,{\rm
ps}^2/{\rm km}$, $\gamma=1.8\times 10^{-3}\,{\rm W}^{-1}\,{\rm
m}^{-1}$, $P_p=0.36$\,W, $c=3\times 10^{8}$\,m/s,
$\lambda_{p1}=1555.92$\,nm, $\lambda_{p2}=1545.95$\,nm, pump FWHM
= 0.8\,nm, signal/idler FWHM = 0.8\,nm.} \label{DTPS2}
\end{figure}

We then generate two curves, from both Eq.~(\ref{CCrate-G}) and
Eq.~(\ref{CCrate-SG}), to fit the experimental data obtained in
the main text. The fitting results are shown in Fig.~\ref{DTPS2}.
As can be seen from the figure, both curves (blue dot-dashed and
purple dotted) fit the experimental data remarkably well. The
super-Gaussian fit appears to give a wider dip width ($\sim
8.0$\,ps FWHM) as compared to the Gaussian fit's result (FWHM dip
width $\sim 6.4$\,ps), which is commensurate with the fact that a
super-Gaussian filter is spectrally narrower (and thus temporally
wider) than its Gaussian counterpart with the same FWHM. A
least-square Gaussian fit to the data (the red solid curve in
Fig.~\ref{DTPS2}), generated by the data-processing program,
suggests that the HOM-dip visibility is $\sim 94.3\%$. It has a
FWHM dip width of about 7.2\,ps, which is right in between the
previous two fitting values. This may be explained by the fact
that the real OBPF employed in the experiment is constructed by a
cascade of a Gaussian filter and a super-Gaussian one, making its
transmission spectrum somewhere in between. In contrast, the two
theoretical fits agree on the ideally attainable dip visibility of
100\%. This result also coincides with the theoretical
understanding from Ref.~\cite{Grice97} that, as long as the
two-photon probability distribution function\footnote{The term is
used loosely here to refers to $F(\nu_s, \nu_i)$ in our
$\chi^{(3)}$ case.} is symmetric with respect to its two frequency
arguments, the HOM dip can achieve a maximum visibility of 100\%.

\begin{figure}
\centerline{\includegraphics[scale=0.6]{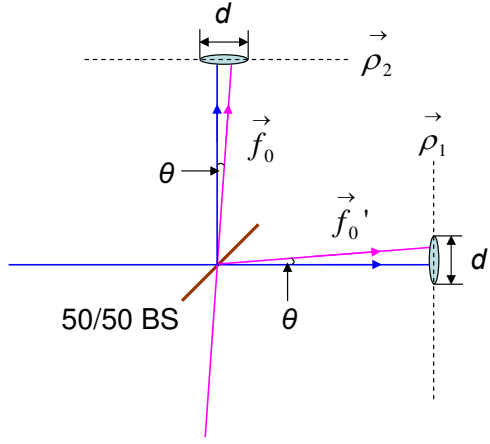}}
\caption{Schematic drawing for investigating one of the possible
scenarios for the less-than-unity HOM-dip visibility: spatial mode
mismatch. See text for details.} \label{SMode}
\end{figure}

There are many reasons that could explain the missing 5.7\%
visibility. For example, one might wonder whether higher-order
dispersion plays a role. But a straightforward Taylor expansion of
the various $k$'s at the central frequency $\Omega$ [similar to
Eq.~(\ref{kvector})] to higher-order terms dismisses this
hypothesis. In fact, the $\Delta k$ quantity is always symmetric
with respect to $\nu_s$ and $\nu_i$ in our $\chi^{(3)}$ two-photon
state due to the isotropic nature of the fiber. However, a little
mismatch between the two OBPFs' spectrum will result in asymmetry
of two arguments in $F(\nu_s, \nu_i)$, and will certainly cause a
degradation of the HOM-dip visibility. Other candidates include:
(i) The real-life BS's performance deviates from an ideal 50/50
BS, i.e., $R+T=1$ and $R \neq T$. This gives rises to a corrective
factor of $\displaystyle\frac{2\,R\,T}{R^2 + T^2}$ to the dip
visibility~\cite{Hong}. When put in the measured values
$(R=0.474,\,T=0.526)$, it gives a 99.4\% corrective coefficient.
(ii) There might be some remaining $\Psi_{2002}$ component due to
the non-ideal alignment of the 50/50 Sagnac loop, which leads to
degradation of the dip visibility~\cite{Halder}. (iii) The
existence of other unsuppressed noise photons, such as Raman
photons and single-pump FWM photons, could also degrade the
attainable dip visibility. (iv) The {\em spatial} modes of the two
photons are not exactly matched at the BS. A simple calculation,
as we will carry out explicitly below, shows that a small angular
mismatch between the two photons' paths distinguishes between the
coincidence-generating amplitudes (TT and RR). As a result, they
are not completely cancelled after the BS, which correspond to the
remaining coincidences at the HOM dip. An angular mismatch as
small as $3\times 10^{-5}$\,rad is required to bring the dip
visibility down to 94.3\%. The details of the calculation go as
follows. Suppose at the BS the two photons intersect at a small
angle $\theta$, as depicted in Fig.~\ref{SMode}. The two
amplitudes can each be written in the Fourier-optic language as:
\bea \psi_{\rm TT} &=& \frac{4}{\pi d^2} e^{i 2\pi \vec{f}_0 \cdot
\vec{\rho}_2}\,, \nonumber\\
\psi_{\rm RR} &=& \frac{4}{\pi d^2} e^{i 2\pi \vec{f}'_0 \cdot
\vec{\rho}_1}\,. \eea Here $d$ is the diameter of the lenses used
to couple light into fibre, $\vec{\rho}_{1,2}$ represent the
two-dimensional coordinates at the lens planes (perpendicular to
the paper), and
$|\vec{f}_0|=|\vec{f}'_0|=\frac{\sin\theta}{\lambda}$ are the
projected wave-vector magnitudes in the lens planes for the two
off-axis waves. The overlap between the two amplitudes, which is
proportional to the HOM-dip visibility~\cite{Rhode}, is given by
\bea \int d\vec{\rho}_1 \int d\vec{\rho}_2\,
\Theta\left(\frac{2\vec{\rho}_1}{d}\right)\,
\Theta\left(\frac{2\vec{\rho}_2}{d}\right)\, \left(\frac{4}{\pi
d}\right)^2 e^{-i 2\pi \vec{f}_0 \cdot \vec{\rho}_2}\, e^{i 2\pi
\vec{f}'_0 \cdot \vec{\rho}_1} &=& \left[\frac{J_1 (\pi d
|\sin\theta| / \lambda)}{\pi d |\sin\theta| / (2
\lambda)}\right]^2 \label{overlap} \eea where $\Theta(\vec{x})$
(it obtains the value 1 when $|\vec{x}|\leq 1$ and everywhere else
zero) represents the effective areas of the lenses, and $J_1(x)$
is the first-order Bessel function. When we put in realistic
values for $d=5$\,mm and $\lambda=1.55$\,$\mu$m, and demand that
Eq.~(\ref{overlap}) yields 0.943, we obtain a numerical solution
for $\theta \simeq \sin\theta \simeq 30\,\mu$rad. From the above
calculation, we can thus see that spatial mode mismatching has the
highest likelihood of contributing to the missing 5.7\% HOM-dip
visibility.

\end{document}